\documentclass[preprint,12pt]{elsarticle}

\usepackage{graphicx}

\begin{document}

\begin{frontmatter}

\title{Electronic structure of non-centrosymmetric superconductor LaPdSi$_3$ and its reference compound LaPdGe$_3$.}

\author{M. J. Winiarski}
\author{M. Samsel-Czeka\l a}
\address{Institute of Low Temperature and Structure Research, Polish Academy of Sciences, Ok\'olna 2, 50-422 Wroc\l aw, Poland}

\begin{abstract}
Electronic structures of a superconductor without inversion symmetry, LaPdSi$_3$, and its non-superconducting counterpart, LaPdGe$_3$, have been calculated employing the full-potential local-orbital method within the density functional theory. The investigations were focused on analyses of densities of states at the Fermi level in comparison with previous experimental heat capacity data and an influence of the antisymmetric spin-orbit coupling on the band structures and Fermi surfaces (FSs) being very similar for both considered here compounds. Their FSs sheets originate from four bands and have a holelike character, but exhibiting pronounced nesting features only for superconducting LaPdSi$_3$. It may explain a relatively strong electron-phonon coupling in the latter system and its lack in non-superconducting LaPdGe$_3$.
\end{abstract}


\journal{Intermetallics}

\end{frontmatter}

\section{Introduction}

The investigations of specific conditions for superconductivity (SC) in non-centrosymmetric compounds were initiated in 2004 y. by the discovery of heavy-fermion SC in CePt$_3$Si with the superconducting critical temperature, T$_c$ = 0.75 K \cite{Bauer_1}. The lack of inversion symmetry of a crystal introduces the so-called antisymmetric spin-orbit coupling (ASOC) that may lead to some unconventional character of SC, i.e., revealing a mixture of spin-singlet and spin-triplet pairings \cite{Rashba, Yuan}.

Among compounds adopting the tetragonal BaNiSn$_3$-type structure \cite{BaNiSn3}, the phenomenon of superconductivity has been reported for LaPdSi$_3$ (T$_c$ = 2.60-2.65 K \cite{Kitagawa, Smidman}), LaPtSi$_3$ (T$_c$ = 1.52 K \cite{Smidman}), LaIrSi$_3$ (T$_c$ = 0.77 K \cite{Okuda}), LaRhSi$_3$ (T$_c$ = 2.16 K \cite{Anand}), BaPtSi$_3$ (T$_c$ = 2.25 K \cite{Bauer_2}), CaPtSi$_3$ (T$_c$ = 2.3 K \cite{Eguchi1}), CaIrSi$_3$ (T$_c$ = 3.3 K \cite{Eguchi2}), SrPdGe$_3$ (Tc = 1.49 K \cite{Miliyanchuk}), and SrPtGe$_3$ (Tc = 1.0 K \cite{Miliyanchuk}). Despite the lack of centrosymmetry in the crystal structure, all these systems exhibit a typically BCS-like SC, except for CaIrSi$_3$ \cite{Eguchi2} having a weakly anisotropic superconducting gap. In turn, in this family, pressure-inducted heavy-fermion SC has been detected in CeIrSi$_3$ (T$_c$ = 1.6 K \cite{Okuda}), CeIrGe$_3$ (T$_c$ = 1.6 K \cite{Honda}), CeRhSi$_3$ (T$_c$ $\approx$ 0.72 \cite{Kimura}), and CeCoGe$_3$ (T$_c$ = 0.7 \cite{Settai}).

The LaPdGe$_3$ compound, recently reported by Tro{\'c} et al. \cite{Troc} as non-superconducting down to 0.3 K, can be considered as an isostructural reference system to the LaPdSi$_3$ superconductor. Interestingly, in the La-Pd-(Ge/Si) intermetallics, forming the layered $Immm$-type structures, the substitution of Si with Ge atoms clearly improves the SC properties \cite{Fujii_1,Fujii_2,Mochiku,Fujii_3,Kasahara}. This effect may be connected with some modifications of specific Fermi surface (FS) features enhancing an electron-phonon coupling \cite{Winiarski}. Therefore, in this paper, electronic structures of LaPdSi$_3$ and LaPdGe$_3$ are probed from first principles to find any differences that might explain an unexpected suppression of SC in LaPdSi$_3$ by the substitution of Ge into Si atomic positions. Analysed are the Sommerfeld coefficients, calculated from densities of states (DOSs) at the Fermi level (E$_F$), $\gamma_e$, and their experimental values of specific heat data,  $\gamma$.
This allows for an estimation of electron-phonon enhancement factors, $\lambda_{ep}$. Furthermore, the FSs topology of the LaPd(Si;Ge)$_3$ is compared to that of other compounds crystallising in the BaNiSn$_3$-type structure as well as other non-centrosymmetric intermetallics, e.g. LaPtSi also exhibiting SC \cite{Kneidinger}.

\section{Computational details}
Band structure calculations have been carried out with the full-potential local-orbital (FPLO-9) code \cite{FPLO}. The Perdew-Wang parametrisation (PW92 \cite{JP}) of the local density approximation (LDA) was applied in both scalar and fully relativistic modes. Experimental values of the lattice parameters of the BaNiSn$_3$-type (space group $I4mm$) unit cell (u.c.), $a$ = 0.4352 nm, $c$ = 0.9661 nm for LaPdSi$_3$ \cite{Kitagawa} and $a$ = 0.45079 nm, $c$ = 0.98478 nm for LaPdGe$_3$ \cite{Troc} were assumed. The atomic (Wyckoff) positions have been fully optimised. Valence-basis sets were automatically selected by the internal procedure of FPLO-9. Total energy values of considered systems were converged with accuracy to ~1 meV for the 16x16x16 {\bf k}-point mesh of the Brillouin zone (BZ), corresponding to 648 {\bf k}-points in the irreducible part of the BZ.

\section{Results and discussion}

The optimised (LDA) atomic positions of LaPdSi$_3$ and LaPdGe$_3$, collected in Table \ref{table1}, only insignificantly differ from the avaible experimental data \cite{Troc}. Interestingly, the accordance of the optimised Wyckoff positions for LaPdSi$_3$ with the experimental values for LaPdGe$_3$ is even better than of those for the Ge-based system itself. This effect is related to the fact that the LDA usually yields an underestimation of the u.c. volume. Therefore, the experimental lattice parameters of LaPdSi$_3$, used in our calculations, are more suitable for the experimental atomic positions of LaPdGe$_3$.

\begin{figure}
\includegraphics[scale=1.0]{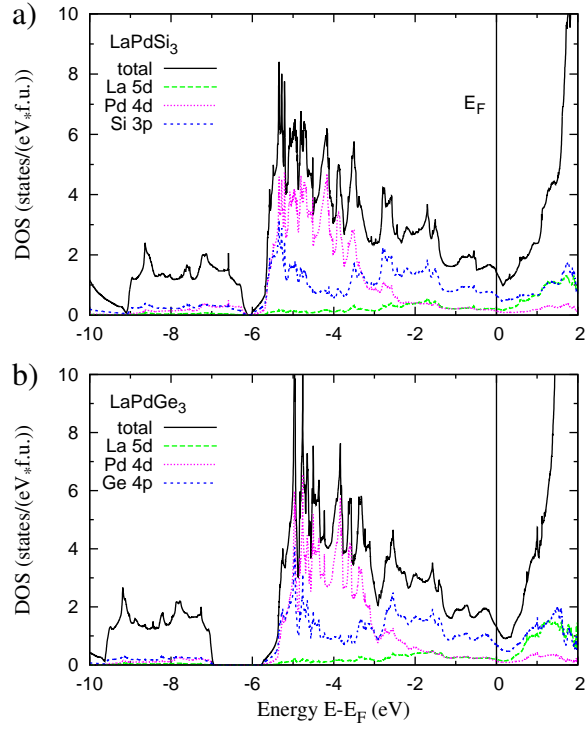}\centering
\caption{Total DOSs of LaPdSi$_3$ and LaPdGe$_3$ and their partial DOS contributions of La, Pd, and Si/Ge atoms, calculated in the fully relativistic approach.}
\label{Fig1}
\end{figure}

The calculated DOSs of considered here systems, presented in Fig. \ref{Fig1}, exhibit a similar shape with narrow peaks of the La 4f states located well above  the Fermi level. The contribution of the La 5d electrons is minor in the whole range of the valence bands below E$_F$. These bands down to -3 eV are dominated by the Si 3p electron orbitals. In turn, the contribution of the Pd 4d electrons forms wide peaks centred at around -4.5 eV and -4.0 eV for the Si- and Ge-based systems, respectively. In LaPdSi$_3$, deep valence bands of the Si 3s electron orbitals (not displayed) are separated by a pseudogap around -6 eV, whereas in LaPdGe$_3$ the Ge 4s electron contribution is located at slightly lower energies (below -7 eV), thus the band-energy gap between -6 eV and -7 eV is completely open. A similar total DOS dependence is also exhibited by other La-Pd-(Si/Ge) intermetallics (La$_3$Pd$_4$(Si;Ge)$_4$ \cite{Winiarski}), but with a less dominant influence of the Si/Ge p states.

The DOS in the vicinity of E$_F$ in both compounds is prevailed by p-electrons coming from Si/Ge atoms like in the case of CaPtSi$_3$ \cite{Bannikov,Kaczkowski}. The calculated DOS at E$_F$, N(E$_F$)=1.42 states/eV/f.u., for LaPdSi$_3$ is only slighlty higher than that obtained for LaPdGe$_3$ (=1.32 states/eV/f.u.), both suggesting a weakly metallic character, being common for intermetallic superconductors and for other members of the BaNiSn$_3$-type family of compounds.

The calculated, $\gamma_e$, and experimental, $\gamma$, values of the Sommerfeld coefficients and T$_c$, as well as electron-phonon enhancement factors, $\lambda_{ep}$, derived from the relation: $\gamma = \gamma_e ( 1 + \lambda_{ep})$, are gathered in Table \ref{table2}. The $\lambda_{ep}= 0.09$ for LaPdGe$_3$ indicates a negligible electron-phonon coupling. Therefore, the BCS-like SC in this compound is impossible. It is worth to note that for LaPdSi$_3$ the obtained $\lambda_{ep}$ values (=0.40-0.59) are comparable to that of BaPtSi$_3$ (0.49) and, generally, adequate for weakly-coupled superconductors.
However, despite the fact that for LaPdSi$_3$ both the T$_c$ and $\gamma_e$ are close to those of CaPtSi$_3$, $\lambda_{ep}$ (=0.21) of the latter compound is significantly lower than that of LaPdSi$_3$ (=0.59). This effect is not observed for CaIrSi$_3$, where the relatively high T$_c$ (=3.3 K) may better correspond to the high $\gamma_e (=4.6$ $mJ/mol/K^2$), but at the same time a resonably moderate value of $\lambda_{ep}$ (=0.43) is obtained. A similar value of $\lambda_{ep}$ (=0.33) has been also reported for LaPtSi with T$_c$ = 3.35 K \cite{Kneidinger}. The low $\lambda_{ep}$ value for CaPtSi$_3$ can be partly explained by worse quality of the polycrystalline samples. 

\begin{figure}
\includegraphics[scale=0.7,angle=-90]{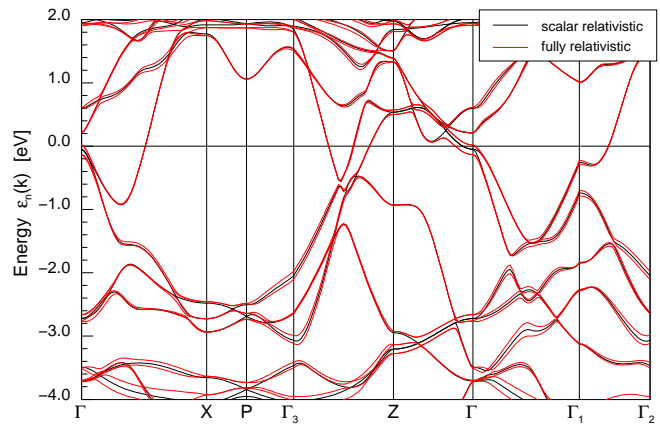}\centering
\caption{Calculated scalar and fully relativistic band structures of LaPdSi$_3$.}
\label{Fig2}
\end{figure}

The influence of an antisymmetric spin-orbit coupling on band structure of LaPdSi$_3$ is presented in Fig. \ref{Fig2}. Although the ASOC-driven splitting is distinct in some regions, e.g. for the conduction bands around the $\Gamma$ point (band splitting is of about 0.2 eV), the overall impact of ASOC on the band structure at $E_F$ and, hence, also the FS of LaPdSi$_3$ is almost negligible. The pronounced ASOC effects are exhibited mainly by valence bands originating from the Pd 4d electrons, yielding relatively large contributions below -3 eV. The fully relativistic calculations for BaPtSi$_3$ \cite{Bauer_2} and CaPtSi$_3$ \cite{Kaczkowski} also revealed only slight changes of the scalar relativistic band structures around E$_F$. Oppositely, in LaPtSi, adopting a different non-centrosymmetric structure of the ThSi$_2$-type, the ASOC-driven FS modifications are significant \cite{Kneidinger}.

\begin{figure}
\includegraphics[scale=1.0]{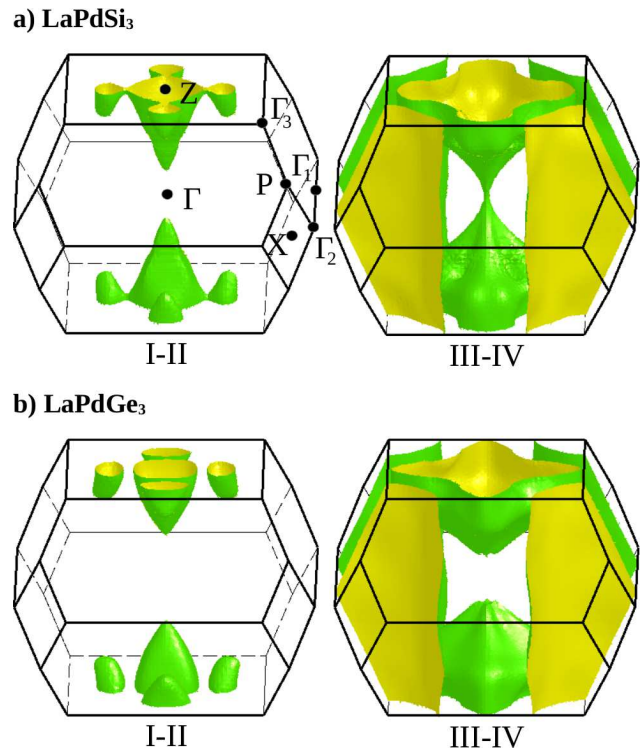}\centering
\caption{Calculated fully relativistic Fermi surfaces of LaPdSi$_3$ and LaPdGe$_3$, containing only holelike sheets. Note that the lift of the Kramers degeneracy of band pairs (I,II) and (III,IV) by the ASOC effects would be not visible in the scale of this figure, hence for each pair only single FS sheet is displayed.}
\label{Fig3}
\end{figure}

The Fermi surfaces of LaPdSi$_3$ and LaPdGe$_3$, consisting of four holelike sheets, calculated in the fully relativistic approach, are depicted in Fig. \ref{Fig3}. Since in both considered systems, the ASOC-driven band splitting is small at $E_F$, the lift of the Kramers degeneracy in pairs of I-II and III-IV FS sheets is also negligible and the sheets within each pair remain almost identical.
Moreover, the differences between the overall shape of the FS sheets of the Si-based superconductor and its Ge-based reference compound seem to be irrelevant. Similarly complex FS sheets to III-IV sheets presented in Fig. \ref{Fig3} have been reported for BaPtSi$_3$ \cite{Bauer_2}, CaPtSi$_3$ \cite{Bannikov,Kaczkowski}, and LaIrSi$_3$ \cite{Okuda}. However, the cylindrical shape of III-IV sheets centered at the X points, is a unique feature of FS exhibited by LaPdSi$_3$ and LaPdGe$_3$. This FS topology is connected with a band gap region around E$_F$, clearly seen in Fig. \ref{Fig2}, along the $X$-$P$-$\Gamma_3$ line.

\begin{figure}
\includegraphics[scale=1.0]{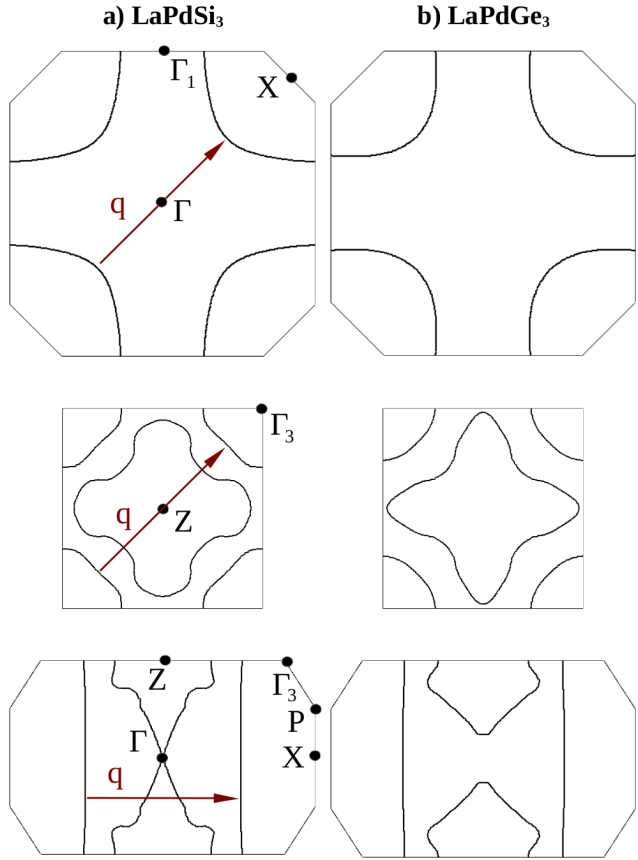}\centering
\caption{Fully relativistic sections of III-IV FS sheets (displayed in Fig. \ref{Fig3}) in LaPdSi$_3$ and LaPdGe$_3$. Possible nesting vector  $\mathbf{q}=(0.5,0.5)\times(2\pi /a)$, spanning surfaces of the holelike cylinders, is marked by red arrows.}
\label{Fig4}
\end{figure}

The sections of III-IV FS sheets for studied here compounds, presented in Fig. \ref{Fig4}, reveal an almost ideal FS nesting with a vector $\mathbf{q}=(0.5,0.5)\times(2\pi /a)$, spanning the cylindrical surfaces. It is worth to underline that for LaPdSi$_3$ this specific nesting vector spans a large area, since the cylinders are distinctly flatter than those of the other system. This feature remains the same for both scalar (not displayed) and fully relativistic FS results. One can suspect that the relatively strong electron-phonon coupling in the Si-based compound may be related to the additional nesting-driven anomaly in phonon dispersion curves. This issue requires a further experimental verification. The analogous idea has also been postulated for other Pd-based superconductor La$_3$Pd$_4$Ge$_4$ \cite{Winiarski}. Furthermore, in the non-centrosymmetric LaPtSi the influence of a similar FS-nesting-driven anomaly on BCS-like SC has been confirmed by calculations of the phonon dispersion curves \cite{Kneidinger}. Meanwhile, such nested FS areas have not been reported for other BaNiSn$_3$-type superconductors, thus this feature is generally not required for an occurrence of the SC in this family of compounds.

\section{Conclusions}
The electronic structures of superconducting LaPdSi$_3$ and its reference compound, LaPdGe$_3$, have been studied from first principles. Although the calculated densities of states of both systems are very similar, the comparison of estimated here Sommerfeld coefficients with those determined from heat capacity data suggests negligible electron-phonon coupling in the non-superconducting LaPdGe$_3$ compound. Moreover, the strength of an antisymmetric spin-orbit coupling in the both studied systems is comparable to that of other compounds crystallising in the BaNiSn$_3$-type structure, but containing no heavy atoms. In turn, the spin-orbit splitting of their bands at the Fermi level is much weaker than that for LaPtSi (different non-centrosymmetric ThSi$_2$-type structure).
Finally, the Fermi surfaces of both investigated compounds exhibit four holelike sheets with unique nesting features, being more pronounced for the superconducting LaPdSi$_3$ than those for its reference Ge-based system. Since in this superconductor the nested area of the calculated FS is substantially larger and its estimated electron-phonon coupling is much stronger than those in the other compound, the latter effect can be caused by a nesting-driven electron-phonon anomaly that could be observed in the phonon dispersion curves. Hence, this issue requires further experimental investigations.

\section*{Acknowledgements}
The authors acknowledge Prof. Ernst Bauer for valuable discussions.
The calculations were performed in Wroc\l aw Centre for Networking and Supercomputing (Project No. 158).

\begin{table}
\caption{Calculated free atomic positions in unit cells of LaPdSi$_3$ and LaPdGe$_3$. Experimental values taken from reference \cite{Troc} are given in parentheses.}
\label{table1}
\begin{center}
\begin{tabular}{llllll} \hline
atom & & x/a & y/b & z/c \\ \hline
LaPdSi$_3$: & & & & \\
La & 2a & 0 & 0 & 0.5978 \\
Pd & 2a & 0 & 0 & 0.2505 \\
Si & 4b & 0 & 1/2 & 0.3625 \\
Si & 2a & 0 & 0 & 0.0000 \\ \hline
LaPdGe$_3$: & & & & \\ 
La & 2a & 0 & 0 & 0.5949 (0.5999) \\
Pd & 2a & 0 & 0 & 0.2480 (0.2496) \\
Ge & 4b & 0 & 1/2 & 0.3587 (0.3548) \\
Ge & 2a & 0 & 0 & 0.0000 (0.0000) \\ \hline
\end{tabular}
\end{center}
\end{table}

\begin{table}
\caption{Experimental, $\gamma$, and calculated, $\gamma_{e}$, values of the Sommerfeld coefficient, derived electron-phonon enhancement factors, $\lambda_{ep}$, and critical temperature values, T$_c$, for LaPdSi$_3$ and LaPdGe$_3$ as well as accessible data for other superconductors adopting the BaNiSn$_3$-type structure.}
\label{table2}
\begin{center}
\begin{tabular}{lllll} \hline
& $\gamma_{e}$ ($\frac{mJ}{mol \cdot K^2}$) & $\gamma$ ($\frac{mJ}{mol \cdot K^2}$) &  $\lambda_{ep}$ &  T$_c$ (K) \\ \hline
LaPdSi$_3$ & 3.34 & 5.3 \cite{Kitagawa} & 0.59 & 2.6 \cite{Kitagawa} \\ 
           &      & 4.67 \cite{Smidman} & 0.40 & 2.65 \cite{Smidman} \\
LaPdGe$_3$ & 3.11 & 3.4 \cite{Troc} & 0.09 & - \\
BaPtSi$_3$ & 3.9 \cite{Bauer_2} & 5.8 \cite{Bauer_2} & 0.49 & 2.25 \cite{Bauer_2} \\
CaPtSi$_3$ & 3.3 \cite{Bannikov,Kaczkowski} & 4.0 \cite{Eguchi1} & 0.21 & 2.3 \cite{Eguchi1} \\
CaIrSi$_3$ & 4.6 \cite{Bannikov,Kaczkowski} & 6.6 \cite{Eguchi2} & 0.43 & 3.3 \cite{Eguchi2} \\ \hline
\end{tabular}
\end{center}
\end{table}

\end{document}